\documentstyle[12pt,epsfig]{article}
\begin{document}
\setlength{\unitlength}{1mm}
\newcommand{\te}{\theta}
\renewcommand{\thefootnote}{\fnsymbol{footnote}}
\newcommand{\bee}{\begin{equation}}
\newcommand{\ene}{\end{equation}}
\newcommand{\bea}{\begin{eqnarray}}
\newcommand{\ena}{\end{eqnarray}}
\newcommand{\dx}{d(x)}
\newcommand{\ux}{u(x)}
\newcommand{\sx}{s(x)}
\newcommand{\tra}{\triangle\theta_}
\newcommand{\uupv}{u^{\uparrow}_{val}}
\newcommand{\udwv}{u^{\downarrow}_{val}}
\newcommand{\dupv}{d^{\uparrow}_{val}}
\newcommand{\ddwv}{d^{\downarrow}_{val}}
\newcommand{\auupv}{\bar{u}^{\uparrow}_{val}}
\newcommand{\audwv}{\bar{u}^{\downarrow}_{val}}
\newcommand{\adupv}{\bar{d}^{\uparrow}_{val}}
\newcommand{\addwv}{\bar{d}^{\downarrow}_{val}}
\newcommand{\uupx}{u^{\uparrow}(x)}
\newcommand{\udwx}{u^{\downarrow}(x)}
\newcommand{\dupx}{d^{\uparrow}(x)}
\newcommand{\ddwx}{d^{\downarrow}(x)}
\newcommand{\auupx}{\bar{u}^{\uparrow}(x)}
\newcommand{\audwx}{\bar{u}^{\downarrow}(x)}
\newcommand{\adupx}{\bar{d}^{\uparrow}(x)}
\newcommand{\addwx}{\bar{d}^{\downarrow}(x)}
\newcommand{\fnx}{{F_{2}^{n}}(x)}
\newcommand{\fpx}{{F_{2}^{p}}(x)}
\newcommand{\aux}{\bar{u}(x)}
\newcommand{\adx}{\bar{d}(x)}
\newcommand{\asx}{\bar{s}(x)}
\newcommand{\gpx}{{g_{1}^{p}}(x)}
\newcommand{\gnx}{{g_{1}^{n}}(x)}

{{\hfill \small \mbox{Universit\'a di Napoli Preprint DSF-34/95 (july 1995)}}}

\begin{center}
{\Large\bf Quantum Statistics and Parton Distributions}
\end{center}
\bigskip\bigskip

\begin{center}
{{\bf G. Miele\footnote{Contributed paper to {\it 17th International
Symposium on Lepton-Photon Interactions}, August 10-15, 1995,
Beijing, China}}}
\end{center}

\bigskip

\noindent
{\it Dipartimento di Scienze Fisiche, Universit\`a di Napoli
''Federico II'', and Istituto Nazionale di Fisica Nucleare, Sezione di Napoli,
Mostra D'Oltremare Pad. 20, I--80125 Napoli, Italy}

\bigskip\bigskip\bigskip

\begin{abstract}
A novel approach to parton distributions parameterization in
terms of quantum statistical functions is here outlined. The
description, already proposed in previous publications, is here
improved by adding to the statistical distributions an
unpolarized {\it liquid} component. This new contribution
to fermion partons is able to reproduce  the expected low $x$
behaviour of structure functions. The analysis provides
a satisfactory description of polarized and unpolarized
deep inelastic data and shows a possible connection between the
Gottfried and Bjorken sum rules.
\end{abstract}

\baselineskip22pt
\newpage

\section{Introduction}
In a recent series of papers \cite{bs1}-\cite{bmmt} the role of Pauli
exclusion principle to explain the experimental data on unpolarized and
polarized structure functions of the nucleons has been studied.

The relevance of the quantum statistics on the parton distribution is
supported by several phenomenological observations. The most relevant
phenomenon is certainly the measurement of a defect in the Gottfried sum
rule \cite{gott, NMC}. It can be explained in terms of a Pauli blocking
effect on the production of $u$ sea quark with respect to $d$ sea quark,
which yields a flavour asymmetry between $\bar{u}$ and $\bar{d}$ in the
proton. Another interesting observation is a relationship which seems to
occur between the shapes of the quark parton distributions and their first
momenta, which is the typical characteristic of Fermi--Dirac distribution
functions.

These considerations naturally suggest to use quantum statistically inspired
parameterizations for the parton distribution functions. In this description,
the independent variable which plays the role of the energy is the Bjorken
variable $x$, and the distributions assumed will be of the Fermi--Dirac kind
for quarks, and Bose--Einstein for gluons. Interestingly, to properly describe
the low $x$ behaviour of the structure functions, a {\it liquid} unpolarized
component dominating the very low $x$ region has to be added. It does not
affect the quark parton model sum rules (QPMSR) but it is necessary to
reproduce the antiquark distribution at low $x$.

The paper is organized as follow, in Section 2 we summarize
the phenomenological motivations behind our description for the parton
distributions. In Section 3 the distribution function parameterizations
are shown in detail and discussed in connection with a possible
physical interpretation. Section 4 deals with the results of a fitting
procedure performed to get the free parameters of the distribution functions.
The theoretical predictions are shown in comparison with the experimental
data and the results for QPMSR are also discussed. In Section 5
we give our conclusions and remarks.

\section{Evidence for quantum statistical effects in parton distributions}

As already stated, an experimental evidence for a central role played by the
Pauli principle in the physics of nucleon is the defect in the Gottfried sum
rule \cite{gott}
\bee
I_{G} = \int_{0}^{1} { 1\over x}\left[\fpx - \fnx \right]~dx = {
1\over 3} ( u + \bar{u} -d -\bar{d} ) ~~~,
\label{9}
\ene
that for $SU(2)_{I}$ invariant sea quark distributions ($\bar{d} = \bar{u}$)
gives $I_G=1/3$. Indeed NMC experiment \cite{NMC} measures for the
l.h.s. of Eq. (\ref{9})
\bee
I_{G} = 0.235 \pm 0.026~~~,
\label{10}
\ene
implying
\bee
\bar{d} - \bar{u} = 0.15 \pm 0.04 ~~~~~~~~~~~~ u - d = 0.85 \pm 0.04~~~.
\label{11}
\ene
This experimental result can be explained following the idea of Field and
Feynman \cite{Field}, who suggested that, in the proton, the Pauli principle
depresses the production of $u \bar{u}$ pairs in the proton with respect to
$d \bar{d}$, since it contains two valence $u$ quarks and only one $d$. We
will return on this fact in the following to focus on a possible connection
between the violation of the above sum rule and the Bjorken one \cite{Bj}.

{}From the previous considerations one expects a relevant role played by the
statistics in the whole phenomenology of deep inelastic scattering, and thus
it suggests to look for others typical characteristics of this behaviour.

A peculiar characteristic of a Fermi--Dirac statistical function is certainly
the strong connection between shape and abundance. This is an immediate
consequence of the Pauli exclusion principle which forbids two or more
fermions to have the same quantum numbers and implies that the more abundant
is distribution the broader is in $x$ the associated function.

With the aim to check if this situation occurs in the nucleon structure
let us consider the abundances of valence quarks in the nucleons. As
it is well-known, at $Q^2 =0$ they are connected to the axial couplings
of the baryon octet $F$ and $D$, through the relations
\bee
\uupv  = 1+F~~~~~~~~~~~~\udwv  = 1-F~~~,
\label{1}
\ene
\bee
\dupv  = {1+F-D \over 2}~~~~~~~~~~\ddwv  = {1-F+D \over 2}~~~.
\label{2}
\ene
By using the experimental values obtained by the two bodies strong decays
of hyperons  $F=0.464 \pm 0.009$ $D=0.793 \pm 0.009$ \cite{pdg94}, \cite{3},
we get
\bee
\uupv \simeq { 3 \over 2} \simeq \udwv + \dupv + \ddwv~~~,
\label{4}
\ene
which tells us that $\uupv$ is the most abundant parton in the proton,
at least at $Q^2=0$. Moreover, by observing that $F \simeq 1/2$ and
$D \simeq 3/4$ one also gets
\bee
\udwv \simeq { 1 \over 2} = {\dupv+\ddwv \over 2 }~~~.
\label{4b}
\ene
In addition to this, the behaviour at high $x$ of $\fnx/\fpx$ \cite{1}, known
since a long time, and the more recent polarization experiments \cite{2},
\cite{2bis}, which show that at high $x$ the partons have spin parallel to
the one of the proton, imply that $\uupx$ is the dominating parton
distribution in the proton at high $x$. Thus, to the most abundant
$u^{\uparrow}$ corresponds effectively a broader distribution in the Bjorken
variable $x$.

Eq. (\ref{4b}) has also another interesting implication. In a previous work
we assumed that the parton distributions at a given large $Q^2$ depend on
their first momenta (abundances) computed at $Q^2 =0$ \cite{bs2}
\bee
p(x) = {\cal F}(x,p_{val})~~~,
\label{5}
\ene
with ${\cal F}$ an increasing function of $p_{val}$ and with a broader shape
for higher values of $p_{val}$. From this assumption and by virtue of
(\ref{4b}) we get
\bee
\udwx = { 1 \over 2} \left[ \dupx + \ddwx \right] = { 1 \over 2}
\dx~~~,
\label{6}
\ene
which implies
\bee
\Delta\ux = \uupx - \udwx = \ux - \dx~~~.
\label{7}
\ene
Note that, Eq. (\ref{7}) connects the contribution of $\Delta\ux$ to
$\gpx$, with the terms due to up and down quarks in the unpolarized
structure functions of nucleons $F_{2}(x)$ \cite{bs2}
\bee
x \gpx \Bigr|_{\Delta u} = {2 \over 3} \left[ \fpx -\fnx
\right]_{u+d}~~~.
\label{8}
\ene
This relation should hold in good approximation for the total
quantities $x \gpx$ and $\fpx -\fnx$, since the contribution in
$\gpx$ due to $\Delta\dx$ is depressed for the twofold reason that
$e_{d}^2 = (1/4) e_{u}^2$ and $\Delta d_{val} \simeq - (1/4)
\Delta u_{val}$. By integrating Eq. (\ref{8}) one thus get a connection
between the spin sum rule and the Gottfried sum rule and in turn a
relation between their possible defects.

\section{Parton distributions as Fermi--Dirac and Bose--Einstein statistical
functions}

The previous considerations on the role played by the Pauli principle
in the nucleon structure suggest to assume Fermi--Dirac distributions
in the variable $x$, at least for large $x$, for the quark partons
\cite{bbmmst}
\bee
p_\lambda(x) = f(x) \left[\exp\left({ x - \tilde{x}(\lambda) \over
\bar{x}} \right) + 1 \right]^{-1}~~~,
\label{px}
\ene
where the index $\lambda$ denotes the different species of quarks,
characterized by flavour and polarization. In Eq. (\ref{px}), $f(x)$ is a
weight function which accounts for the energy level density, and because it
is connected to the nonperturbative aspect of QCD results independent of
flavours and polarization. The {\it universal} parameter $\bar{x}$ represents
the {\it temperature} for the system, whereas $\tilde{x}(\lambda)$ stands the
{\it thermodynamical potential} of the parton $\lambda$.

The expression chosen for $f(x)$ is inspired by the expected power behaviour
at $x=0$, and by the obvious kinematical cut
which forces the function to vanish at $x=1$. In order to satisfy
these constraints we assume for simplicity a power low dependence on $x$
\bee
f(x)= A~ x^{\alpha} (1 -x)^{\beta}~~~.
\label{fx}
\ene
Indeed, the assumption that the form given in Eq. (\ref{px}) for the quark
distribution functions, which requires different thermodynamical potentials
in order to describe the experimental data, is valid in the low $x$
limit as well has at least two unpleasant features. Firstly, one gets in the
nonperturbative region different behaviour for the different parton
distributions, where on the contrary one would expect an {\it universal}
dependence on $x$. Moreover, the power dependence on $x$ of Eq. (\ref{fx}),
fitted by the experimental data mostly placed at the large $x$, is not
suitable to reproduce the more divergent contribution expected at low $x$.
This most divergent part does not contribute to QPMSR as the ones given by
Gottfried and Bjorken \cite{Bj} with $I=1$ quantum numbers exchanged.
To this aim we add to (\ref{px}) an unpolarized component, which we call
{\it liquid} to stress the possibility that it is connected to the presence,
at low $x$, of a new phase in the quark-gluon plasma due to the highly
nonperturbative QCD regime
\bee
p_\lambda(x) = {A_{L} \over 2}~x^{\alpha_{L}} (1-x)^{\beta_L} +
A~ x^{\alpha} (1 -x)^{\beta} \left[\exp\left({ x - \tilde{x}(\lambda)
\over \bar{x}} \right) + 1 \right]^{-1}~~~.
\label{ptx}
\ene
In the fitting procedure we take as free parameters, apart from the constants
involved in $f(x)$ and in the {\it liquid} component of (\ref{ptx}), the
temperature $\bar{x}$ and the $\tilde{x}$ for $u^{\uparrow(\downarrow)}$,
$d^{\uparrow(\downarrow)}$, $\bar{u}$ and $\bar{d}$ (the latters are assumed
not polarized). We tried initially to introduce spin-dependent $\tilde{x}$'s
also for the $\bar{q}$'s and to test the relationship
\bee
\Delta \bar{u}(x) = \bar{u}(x) - \bar{d}(x)~~~,
\label{1a}
\ene
assumed in previous works \cite{bs2, bos1}, but
unfortunately, we found practically the same $\chi^2$ with negative
and positive values for $\Delta\bar{u}(x)$ and/or $\Delta\bar{d}(x)$.
Hence, for not loosing predictivity in the fit procedure we have
assumed unpolarized antiquarks.

As far as the strange quarks are concerned, we assume for simplicity
unpolarized distribution functions given by the empirical expression
\bee
s = \bar{s} = {\bar{u}+\bar{d} \over 4}~~~,
\label{sx}
\ene
which experimentally is very well satisfied.

Analogously, for the gluons, if we neglect their polarization , the
bosonic statistic suggests the consider, at large $x$, the distribution
function
\bee
G(x) = {16 \over 3} A~ x^{\alpha} (1 -x)^{\beta}
\left[\exp\left({ x - \tilde{x}_{G} \over \bar{x}} \right) - 1
\right]^{-1}~~~,
\label{gx}
\ene
where the factor $16/3$ is just the product of $2$ ($S_{z}(G) = \pm 1$)
times $8/3$, the ratio of the colour degeneracies for gluons and
quarks.

\section{Discussion of the results}

By assuming for the parton distributions Eqs. (\ref{ptx}) and (\ref{gx}),
we fit the distribution parameters from the experimental data for
$\fpx -\fnx$ \cite{NMC}, $xF_{3}(x)$ \cite{12}, $x\gpx$ \cite{2, 2bis} and
$x\gnx$ \cite{16}, which do not receive contributions from the {\it liquid}
component, and from $\fnx/\fpx$ \cite{NMC} and $x\bar{q}(x)$ \cite{9}.

The avaliable experimental data on deep inelastic scattering observables
correspond in general to different values of $Q^2$. This would suggest,
in order to use the data to determine the distribution parameters, which in
general will depend on $Q^2$, to apply the evolution equations to lead all
the experimental results to the same $Q^2$. In our analysis we have neglected
this $Q^2$ dependence of the distribution parameters, since we expect from
the evolution equations a smooth logarithmic dependence.
As far as the polarized
distributions is concerned, in fact, the expected $Q^2$ dependance \cite{bos1}
results to be smaller than the experimental errors, and thus our analysis
is slightly affected by neglecting this dependance.

Indeed, the data on unpolarized nucleons structure functions are at
$Q^2 = 4~GeV^2$ \cite{NMC}, the neutrino data at $Q^2 = 3~GeV^2$ \cite{12},
and $\bar{q}$ measures are performed at $Q^2 = 3~GeV^2$ and $5~GeV^2$
\cite{9} and differ at small $x$, while our curve is intermediate
between the two sets of
data. The data on $g_1^n(x)$ are at $Q^2 = 2~GeV^2$ \cite{16},
whereas $g_1^p(x)$ is measured at $Q^2 = 10~GeV^2$ by SMC \cite{2} and
at $Q^2 = 3~GeV^2$ by E143 \cite{2bis}; despite some narrowing of
the distribution at higher $Q^2$ showing up in the data, the values of
$I_p$ are in good agreement.

In Table 1 we report the parameters found in the present analysis
\cite{bmmt}and compare them with the results of a previous fit
(without liquid) \cite{bbmmst}, and with the ones by Bourrely and
Soffer \cite{bos1} found on similar principles, but with several different
assumptions. In the Figures 1.-6., the predictions for the nucleon structure
$\fpx -\fnx$, $\fnx/\fpx$, $x\gpx$, $x\gnx$, $xF_{3}(x)$ and $x\bar{q}(x)$
are shown, respectively, and compared with the experimental data.

The parton distributions found in \cite{bmmt} are described in Figure 7.
Since the total
momentum carried by fermion partons is $53\%$, we get $\tilde{x}_G=-1/15$
by requiring that the gluons carry out the remaining part of the
proton momentum. In Ref. \cite{bmmt},
the gluon distribution is compared with the information
found on them in CDHSW \cite{CDHSW}, SLAC$+$BCDMS \cite{SLAC&BCDMS} and
in NMC \cite{NMCgluon} experiments at $Q^2 = 20~GeV^2$. The
agreement is fair for $x > .1$, while the fast increase at small $x$,
confirmed also from the data at very small $x$ at Hera \cite{H1gluon},
confirms that a liquid component is needed also for gluons. The excess
at high $x$ of our curve with respect to experiment may be, at least
in part, explained by the expected narrowing of the distribution from
$Q^2 = 4~GeV^2$, where we fit the unpolarized distributions,
to $Q^2 = 20~GeV^2$.

The inclusion of the {\it liquid} term and the extension of our fit to
the precise experimental results on neutrinos has brought to substantial
changes in the parameters \cite{bmmt} with respect to the previous work
\cite{bbmmst}.

The low $x$ behaviour of $f(x)$ become smoother ($\simeq x^{-.203 \pm .013
}$ instead of $x^{-0.85}$), but this is easily understood since the
previous behaviour was a compromise between the smooth {\it gas}
component and the rapidly changing {\it liquid} one to reproduce the
behaviour of $\bar{q}(x)$. The {\it liquid} component, relevant only
at small $x$, carries only $.6\%$ of parton momentum and its behaviour
$\sim x^{-1.19}$, similar to the result  found in \cite{capella},
is less singular than the one, suggested in the
framework of the multipherial approach to deep-inelastic scattering,
proportional to $\sim x^{-1.5}$ \cite{15}. The parameter
$\tilde{x}(u^{\uparrow})$ took the highest value allowed by us (1.),
since the factor in $f(x)$,
$(1-x)^{2.34}$, is taking care to decrease $\uupx$ at high $x$. The
temperature $\bar{x}$ is  larger than
the previous one and the one found by Bourrely and Soffer \cite{bos1}. Instead
$\tilde{x}(u^{\downarrow})$ is slightly smaller than the previous
determination \cite{bbmmst} and about half the value found in \cite{bos1},
where $f(x)$ is different for $u^{\uparrow}$ and $u^{\downarrow}$.

The ratio $r = \udwx/\dx$ varies in the
narrow range $(.546,.564)$ in fair agreement with the constant value
$1 -F = .536 \pm .009$ assumed in \cite{bbmmst} and
slightly larger than the value $1/2$ taken in \cite{bs2} and \cite{bos1}.

The central value found for the first moment of $\bar{u}_{gas}(x)$, $.03$,
is smaller than $\bar{d}_{gas}(x)/2$, $.08$, while Eq. (\ref{1a})
implies $\bar{u}(x) \geq \bar{d}(x)/2$. However, the large upper error on
$\bar{u}_{gas}$ and the uncertainty
in disentangling the gas and liquid contributions for the $\bar{q}$'s do not
allow to reach a definite conclusion about the validity of Eq. (\ref{1a}).

Finally, we can compare the predictions of \cite{bmmt}
with the measured asymmetry for Drell-Yan
production of muons at $y=0$ in $pp$ and $pn$ reactions
\bee
A_{DY} = { d \sigma_{pp}/dy - d \sigma_{pn}/dy \over
d \sigma_{pp}/dy + d \sigma_{pn}/dy}~~~,
\label{ady}
\ene
which at rapidity $y=0$ reads
\bee
A_{DY} = { (\lambda_s(x) -1) (4 \lambda(x) -1) +
(\lambda(x) -1) (4 \lambda_s(x) -1) \over
(\lambda_s(x) +1) (4 \lambda(x) +1) +
(\lambda(x) +1) (4 \lambda_s(x) +1) }~~~,
\label{ady1}
\ene
where $\lambda_s(x)=\bar{u}(x)/\bar{d}(x)$ and
$\lambda(x)=u(x)/d(x)$. At $x=.18$ we have $\lambda_s(.18)=.454$
and $\lambda(.18)=1.748$ giving rise to $A_{DY}(.18)=-.138$ in
fair agreement with the experimental result $-.09 \pm .02 \pm .025$
\cite{NA51}.\\
The behaviour of $A_{DY}(x)$ is plotted in Figure 8 together with
the experimental point measured by NA51 collaboration.

\section{Conclusions}

We compared with data the quark-parton distributions given by the sum of
Fermi--Dirac functions and of a term not contributing to the QPMSR relevant at
small $x$. We obtain a fair description for the unpolarized and polarized
structure functions of the nucleons as well as for the $F_3(x)$ precisely
measured in (anti)neutrino induced deep-inelastic reactions and for $\bar{q}$
total distribution. The conjectures of previous works on $d$ distributions are
well confirmed by the values chosen for their thermodynamical potentials. As
long as the implications for QPMSR the values found for the first momenta of
the various parton species give l.h.s.'s consistent with experiment. For the
fundamental issue of the Bjorken sum rule, as advocated in previous works
\cite{bs1, bs4} and \cite{bbmmst}, we get
\bea
\Delta u & \simeq  & u - d + 2F-1~~~,\label{37}\\
\Delta d & \geq  & F- D~~~,\label{38}
\ena
to confirm the suspicion of a violation of Bjorken sum rule related to
the defect in the Gottfried sum rule.

A word of caution is welcome for our conclusions on the violation of
Bj sum rule, since we did not include the effect of QCD corrections
in relating the quark parton distributions to the structure functions.
Also we assumed no polarization for $\bar{q}$, being unable to get a
reliable evaluation of $\Delta \bar{q}$ with the present precision
for the polarized structure functions at
small $x$. Indeed our description of $g_1^p(x)$ and $g_1^n(x)$
is good in terms of $\Delta u(x)$ and $\Delta d(x)$, but our
prediction is smaller than the central values of the three lowest $x$
values measured by SMC.

\newpage
\bigskip\bigskip
\par\noindent
\begin{tabular}{|c|c|c|c|c|}
\hline
Parameters
& Previous fit  \cite{bbmmst} & Fit BS \cite{bos1}&\multicolumn{2}{c|}
{Actual fit \cite{bmmt}}\\
& & & \multicolumn{2}{c|}{$\chi^2/N=2.47$}\\
\hline
$A$ & $.58$ &  & \multicolumn{2}{c|}{$ 2.66\begin{array}{c} +.09\\ -.08
\end{array} $}\\
$\alpha$ & $-.85$ & $\begin{array}{c} -.646~\mbox{for}~u^{\uparrow}_{val}\\
-.262~\mbox{for}~u^{\downarrow}_{val}\end{array} $ &
\multicolumn{2}{c|}{$ -.203 \pm .013$}\\
$\beta$ &  &  &  \multicolumn{2}{c|}{$ 2.34
\begin{array}{c} +.05\\ -.06 \end{array}$}\\
$A_{L}$ &  &  & \multicolumn{2}{c|}{$ .0895 \begin{array}{c} +.0107\\
-.0084\end{array}$}\\
$\alpha_{L}$ & & & \multicolumn{2}{c|}{$ -1.19 \pm .02$}\\
$\beta_{L}$ & & & \multicolumn{2}{c|}{$ 7.66 \begin{array}{c} +1.82\\
-1.59\end{array}$}\\
\cline{4-5}
$\bar{x}$& $.132$ & $.092$ & $.235 \pm .009$ & gas abund. \\
\cline{5-5}
$\tilde{x}(u^{\uparrow})$& $.524$&  $.510~\mbox{for}~u^{\uparrow}_{val}$
& $1.00 \pm .07$ & $1.15 \pm .01$\\
$\tilde{x}(u^{\downarrow})$& $.143$ & $.231
{}~\mbox{for}~u^{\downarrow}_{val}$ & $.123 \pm .012$ &
$.53 \pm .01$\\
$\tilde{x}(d^{\uparrow})$&  & & $-.068 \begin{array}{c} +.021\\
-.024\end{array}$ & $.33 \pm .03$\\
$\tilde{x}(d^{\downarrow})$&  &  &$.200 \begin{array}{c} +.013\\
-.014\end{array}$ & $.62 \pm .01$\\
$\tilde{x}(\bar{u}^{\uparrow})
$& $-.216$ &  & $-.886 \pm .266$ & $.015 \begin{array}{c} +.034\\
-.009\end{array}$\\
$\tilde{x}(\bar{u}^{\downarrow})$& $-.141$ &  & $''$ & $''$\\
$\tilde{x}(\bar{d}^{\uparrow})=\tilde{x}(\bar{d}^{\downarrow})
$& $''$ &  & $-.460\begin{array}{c} +.047\\
-.064\end{array}$ &$.08 \begin{array}{c} +.03\\
-.02\end{array}$\\
\hline
\end{tabular}\\
\footnotesize
{\bf Table 1.} Comparison of the values for the parameters of
our best fit \cite{bmmt} with the corresponding\\ quantities, if any, found
in previous analysis \cite{bbmmst}, \cite{bos1}.\\
\normalsize

\bigskip\bigskip\bigskip

\par\noindent
\begin{tabular}{|c|c|c|c|}
\hline
Sum rule & Experimental data & Our fit \cite{bmmt}& QPM \\
\hline
GLS & $2.50 \pm .018 \pm .078$ \cite{12}
& $2.44\begin{array}{c}+.04\\-.07\end{array}$ & $3$ \\
G & $.235 \pm .026$ \cite{NMC}
&$.20 \pm .02$ &$1/3$ \\
EJ $\left\{\begin{array}{c} I_{p}\\ ~ \\ ~ \\ I_{n} \end{array}\right.$ &
$\begin{array}{c} .136 \pm .011 \pm .011~ \cite{2}
\\ .129 \pm .004 \pm .009~ \cite{2bis}
\\ \\ -.022 \pm .011~ \cite{16} \end{array}$ &
$\begin{array}{c} {.122 \pm .007}\\
{-.030 \pm .010}\end{array}$ &
$\begin{array}{c} .188 \pm .005\\ ~ \\-.021 \pm .005\end{array}$ \\
Bj & & $.152 \pm .010$ & $.209$\\
\hline
\end{tabular}\\
\footnotesize
{\bf Table 2.} Comparison of our predictions for the sum rules
with the experimental\\ values and with the quark parton model (QPM)
predictions without QCD corrections.
\normalsize
\newpage
\begin{figure}

\begin{center}
\leavevmode
\epsfxsize=170mm \epsfbox{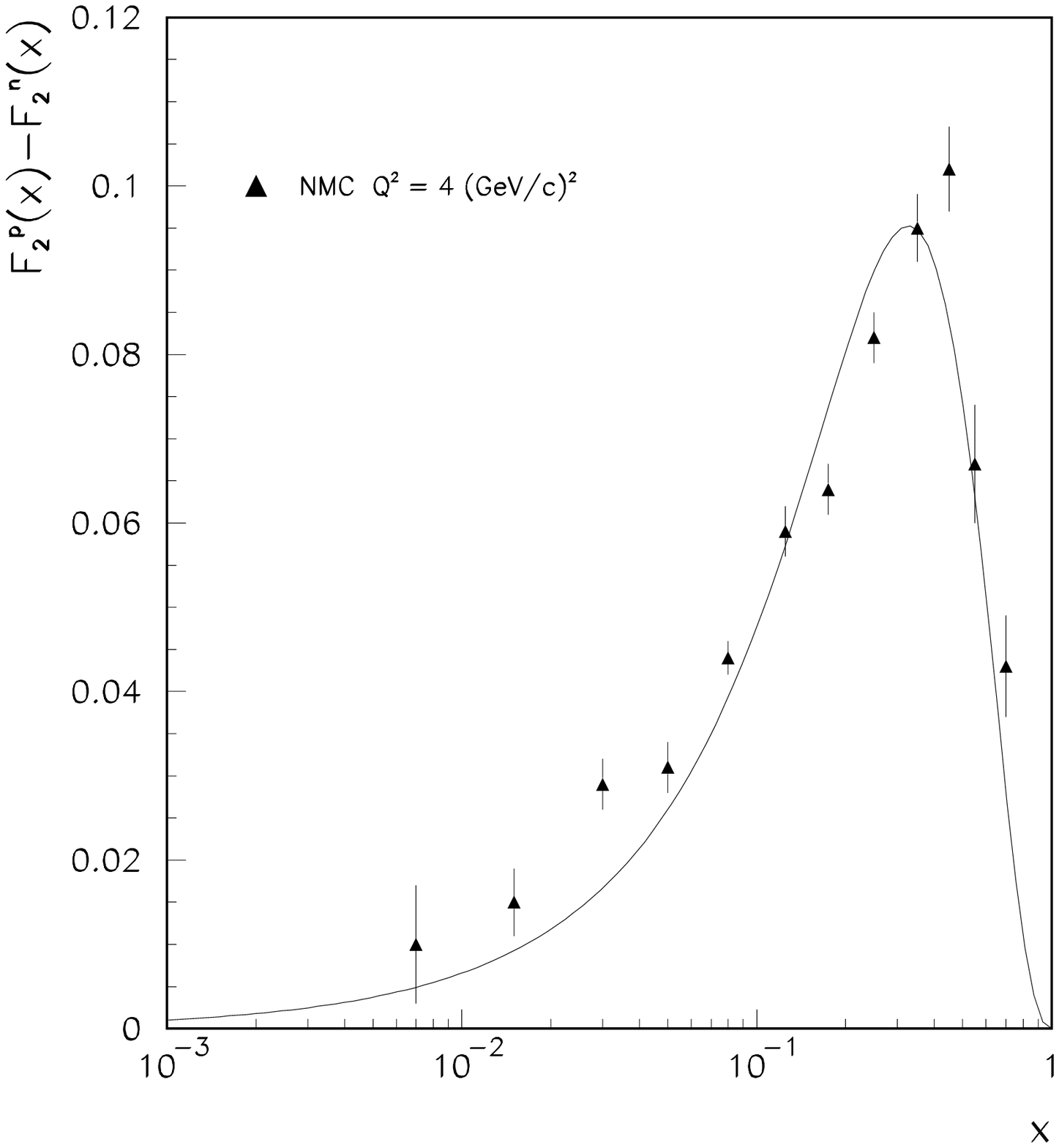}
\end{center}
\caption{The prediction for $\fpx-\fnx$ is plotted and compared with
the experimental data [10]}
\label{fig1}
\end{figure}
\newpage
\begin{figure}

\begin{center}
\leavevmode
\epsfxsize=170mm \epsfbox{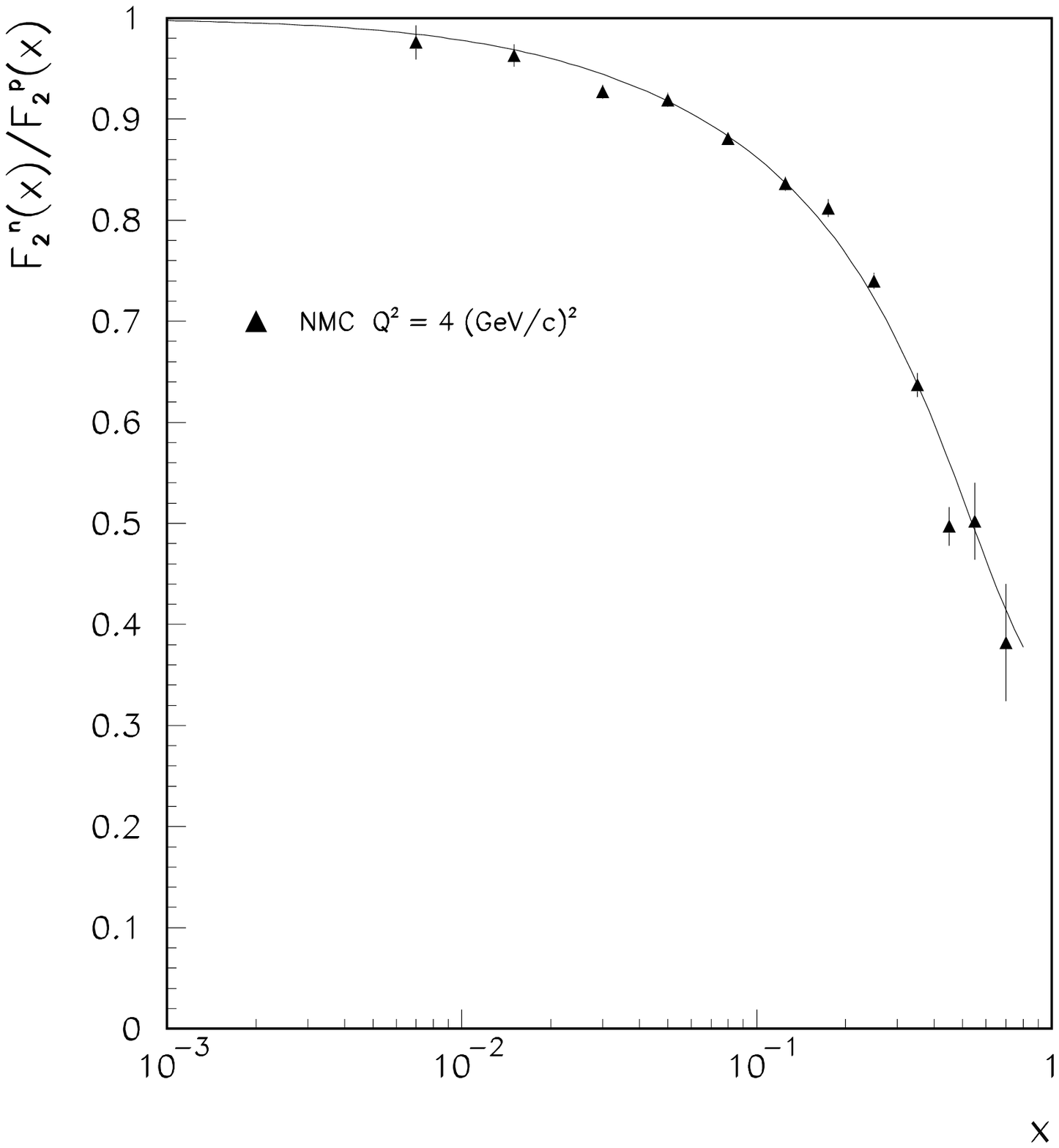}
\end{center}
\caption{The prediction for $\fnx/\fpx$ is plotted and compared with
the experimental data [10].}
\end{figure}
\newpage
\begin{figure}

\begin{center}
\leavevmode
\epsfxsize=170mm \epsfbox{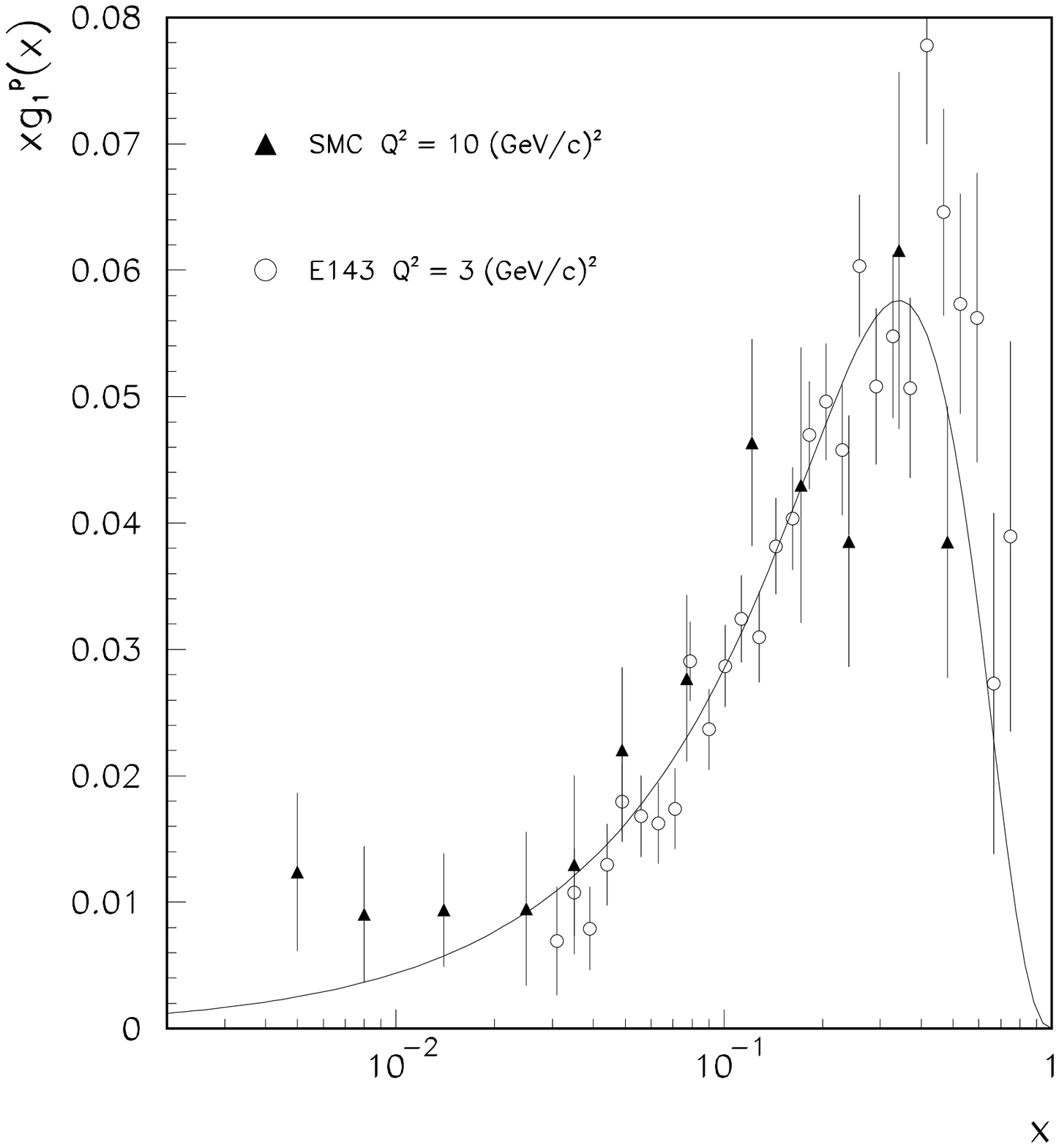}
\end{center}
\caption{$x\gpx$ is plotted and compared with the data [18], [19].}
\end{figure}
\newpage
\begin{figure}

\begin{center}
\leavevmode
\epsfxsize=170mm \epsfbox{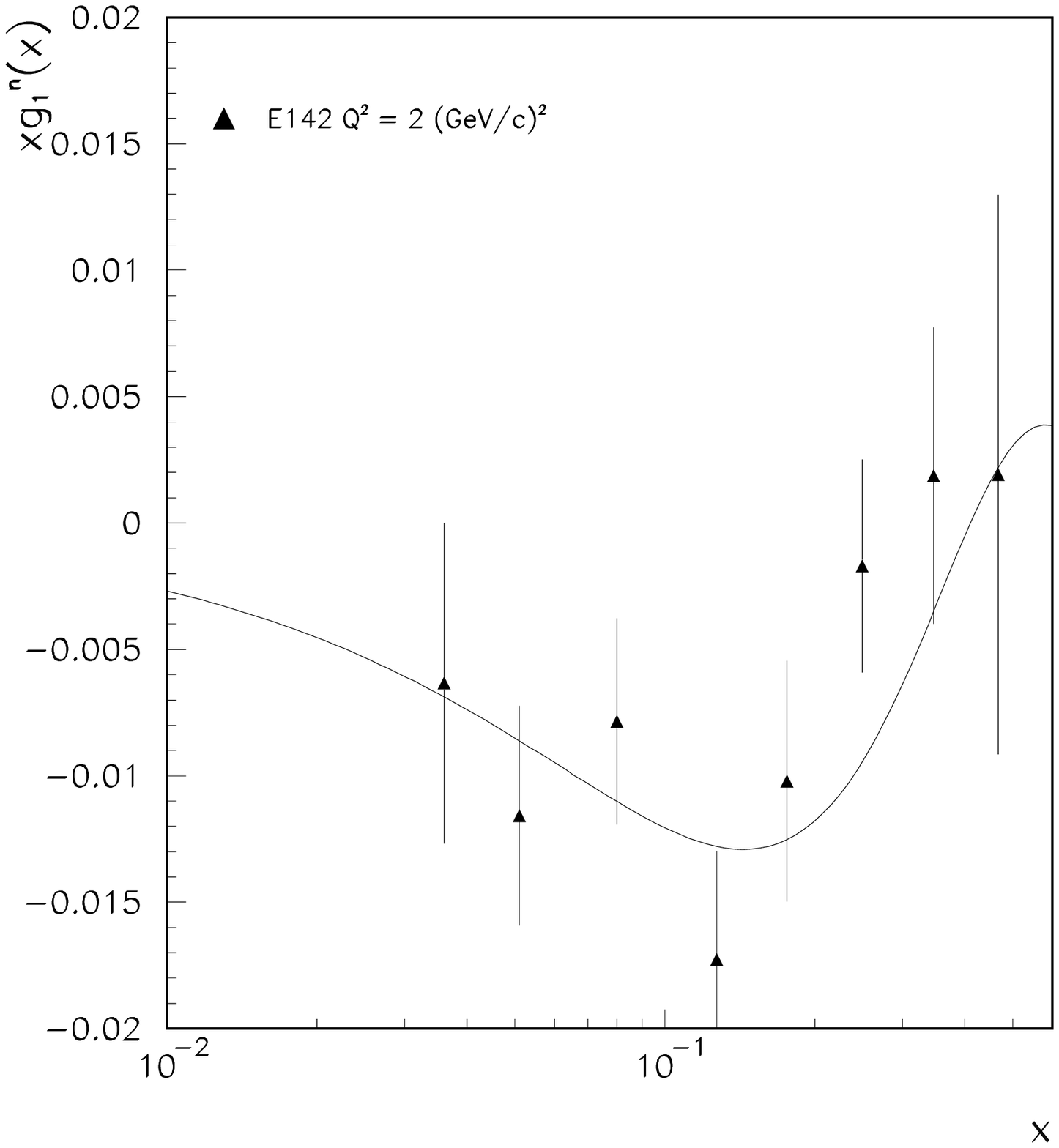}
\end{center}
\caption{$x\gnx$ is plotted and compared with the data [21].}
\end{figure}
\newpage
\begin{figure}

\begin{center}
\leavevmode
\epsfxsize=170mm \epsfbox{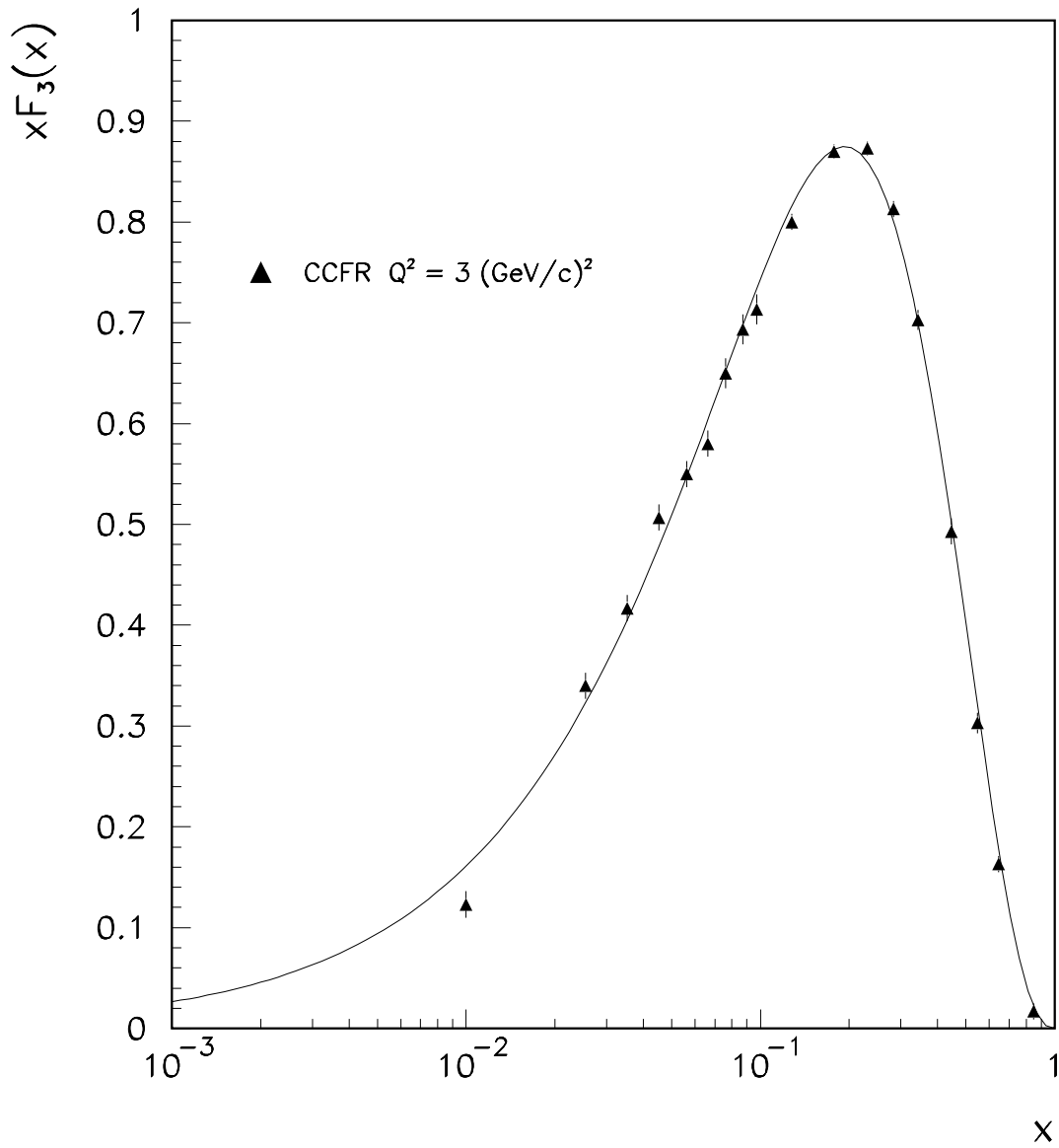}
\end{center}
\caption{$xF_{3}(x)$ is plotted and the experimental values are taken
from [20].}
\end{figure}
\newpage
\begin{figure}

\begin{center}
\leavevmode
\epsfxsize=170mm \epsfbox{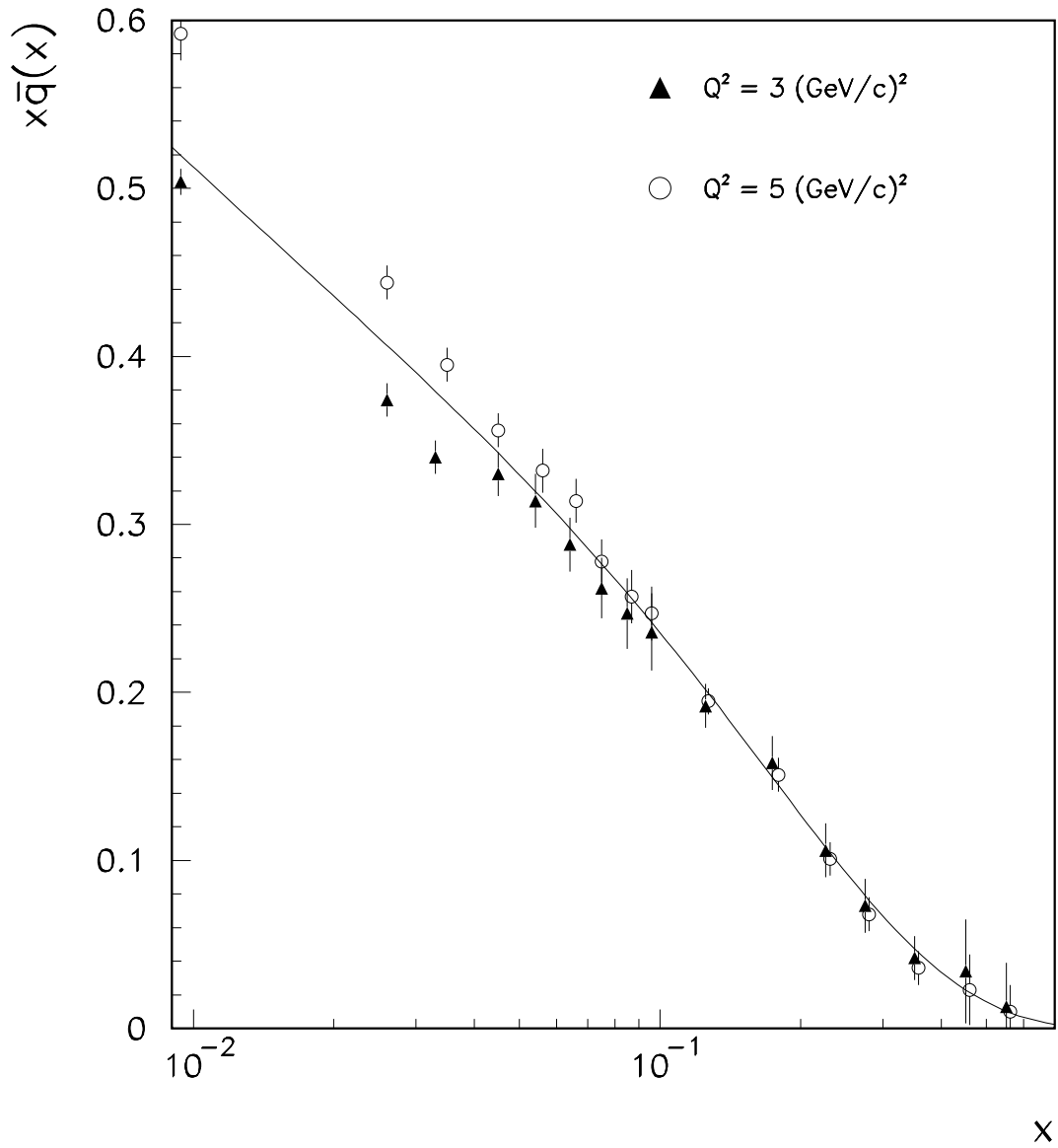}
\end{center}
\caption{$x\bar{q}(x)$ versus $x$ is shown, the experimental data
correspond to [22].}
\end{figure}
\newpage
\begin{figure}

\begin{center}
\leavevmode
\epsfxsize=170mm \epsfbox{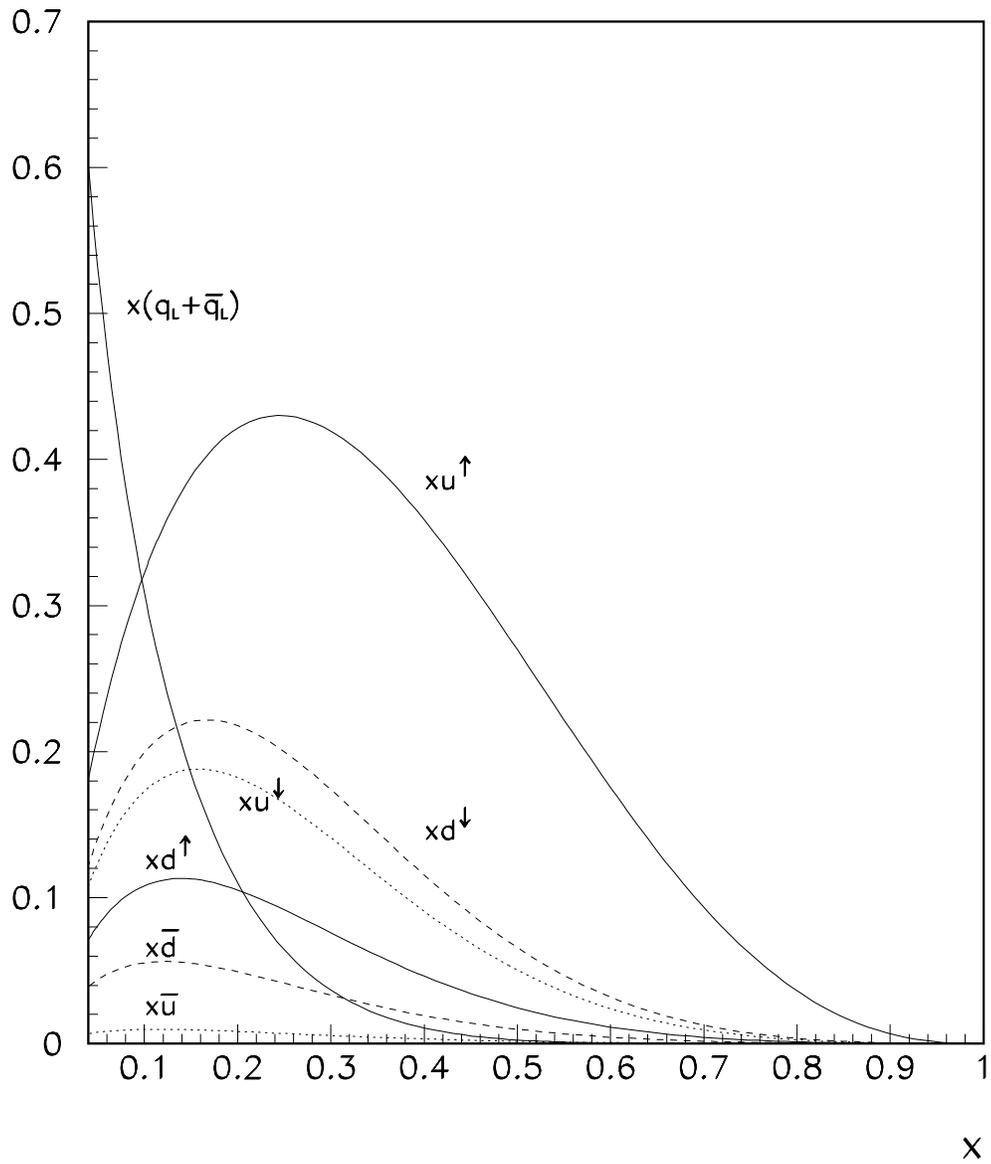}
\end{center}
\caption{The momentum distributions of {\it gas} component of
$q$ and $\bar{q}$'s, and of the total {\it liquid} part are here shown.}
\end{figure}
\newpage
\begin{figure}

\begin{center}
\leavevmode
\epsfxsize=170mm \epsfbox{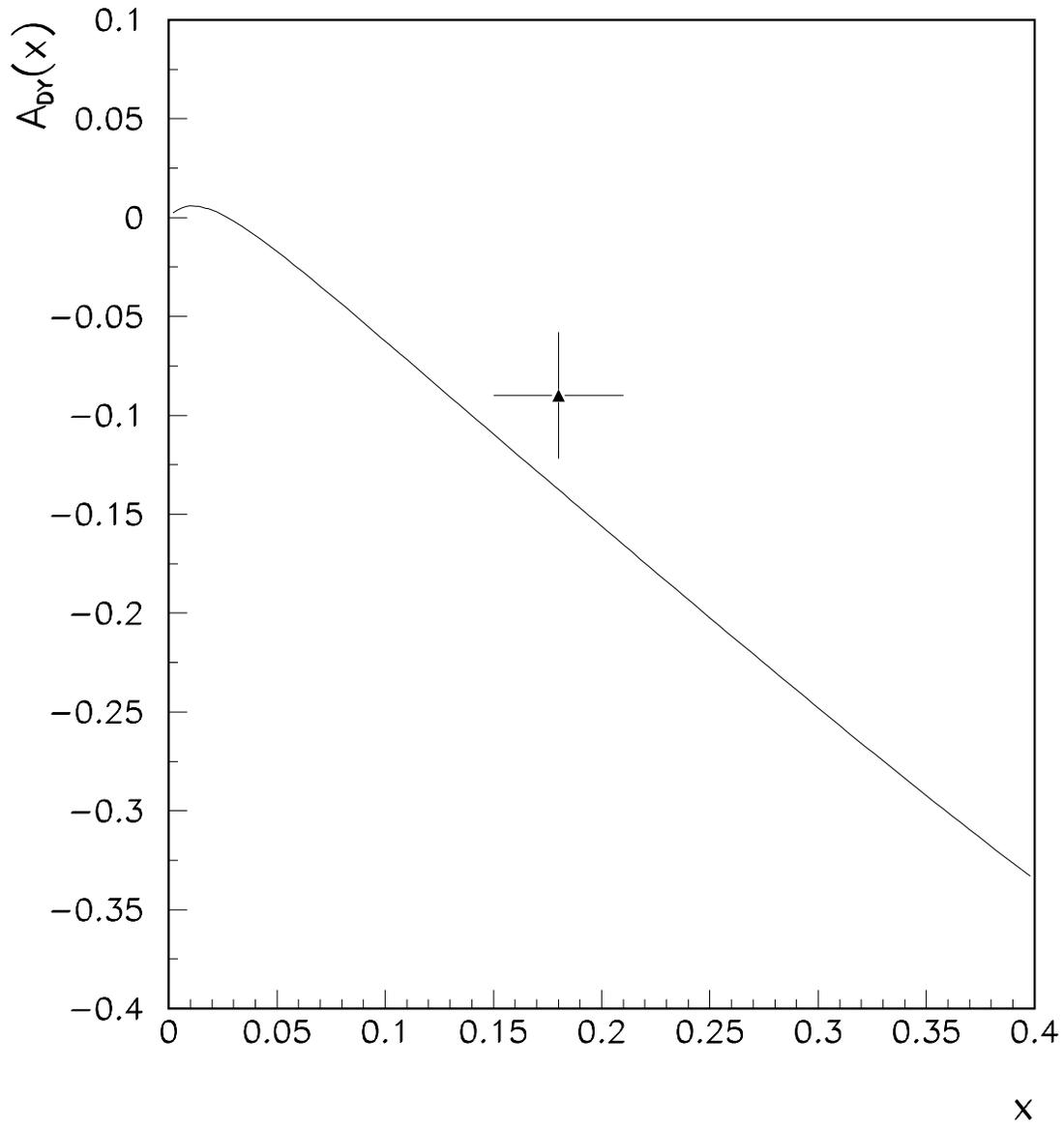}
\end{center}
\caption{The asymmetry $A_{DY}(x)$ is here plotted,
the experimental result is taken from [29].}
\end{figure}
\end{document}